\documentclass[journal]{vgtc}                     

\onlineid{1487}

\vgtccategory{Research}

\vgtcpapertype{technique}

\title{A Deixis-Centered Approach for Documenting Remote Synchronous Communication around Data Visualizations}

\author{%
  Chang Han 
  and
  Katherine E. Isaacs
}

\authorfooter{
  \item Chang Han and Katherine E. Isaacs are with the University of Utah. E-mail: \{changhan, kisaacs\}@sci.utah.edu.
}

\abstract{
Referential gestures, or as termed in linguistics, {\em deixis}, are an essential part of communication around data visualizations. Despite their importance, such gestures are often overlooked when documenting data analysis meetings. Transcripts, for instance, fail to capture gestures, and video recordings may not adequately capture or emphasize them. We introduce a novel method for documenting collaborative data meetings that treats deixis as a first-class citizen. Our proposed framework captures cursor-based gestural data along with audio and converts them into interactive documents. The framework leverages a large language model to identify word correspondences with gestures. These identified references are used to create context-based annotations in the resulting interactive document. We assess the effectiveness of our proposed method through a user study, finding that participants preferred our automated interactive documentation over recordings, transcripts, and manual note-taking. Furthermore, we derive a preliminary taxonomy of cursor-based deictic gestures from participant actions during the study. This taxonomy offers further opportunities for better utilizing cursor-based deixis in collaborative data analysis scenarios.
}


\teaser{
  \centering
 \includegraphics[width=\linewidth, alt={
 The image presents an overview of a proposed framework for generating interactive meeting documents. The framework consists of four stages: Data Collection (A), Utterance Matching (B), Reference Extraction (C), and Interactive Notes Generation (D). In the Data Collection stage, there are icons showing two participants, Alice and Bob, are having a remote meeting. They are in the middle of a meeting, there is also a clock showing the current time: 07:21.071. There is a scatter plot below Alice and Bob, representing what they are looking at through a shared screen. There are two laser pointers on top of the scatter plot, each represents one of the participant. Alice said "Look at this cluster here", and drew a circle on the scatter plot using her laser pointer at the same, leaving an orange trail of a circle. Then through matching transcripts with gestures, the framework goes to the second stage, utterance matching. In this stage, the framework matches utterances with referential gestures made by the participants. It associates the spoken words from the transcripts with specific gestures, such as pointing or highlighting areas on a data visualization. On the figure there four examples. Example 1: Bob's utterance "The nodes below this line are..." is matched with a gesture where Bob uses a laser pointer to draw a line beneath certain nodes in the scatter plot. Example 2: Bob's utterance "this node over here could be..." is matched with a gesture pointing to a specific node by drawing an arrow. Example 3: Alice's utterance "This area looks..." is paired with her gesture doing a Z scan on a particular cluster. Example 4: Alice's utterance "Look at this cluster here. I think it is..." is linked to her circling a highlighted cluster. Then, through extracting word references, the framework goes to the third stage, reference extraction. In this stage, the framework goes a step further by extracting detailed references from the matched utterances and gestures. This involves identifying specific words and phrases that refer to gestures. Example 1. Bob'utterance, "The nodes below this line are...", the system identifies "nodes below this line" as a reference to a specific subset of data points. Similarly, "this node over here could be..." extracts "this node" as a reference to a specific data point. Alice's references to "This area" and "this cluster" are also extracted. Then, through generating interactive meeting document, the frame work goes to the final stage, interactive notes. This stage presents an interactive meeting document that incorporates the processed information from the previous stages. The texts are on the left, while the chart used during the conference is placed on the right. This document includes automatically generated transcripts and meeting minutes, augmented with visual annotations that reflect the references made during the meeting. For instance, the meeting minutes might summarize Alice's observation as "Alice finds a cluster that...", and the transcripts will show detailed utterances like "(07:21.071) Alice: Look at this cluster here. I think it is..." highlighted in the text. There is a hand cursor pointing at the highlighted words "this cluster". At the same time, a circle around the certain cluster shows up in the scatter plot, with a small "Alice" label around the circle, indicating that this is made by Alice. 
 }]{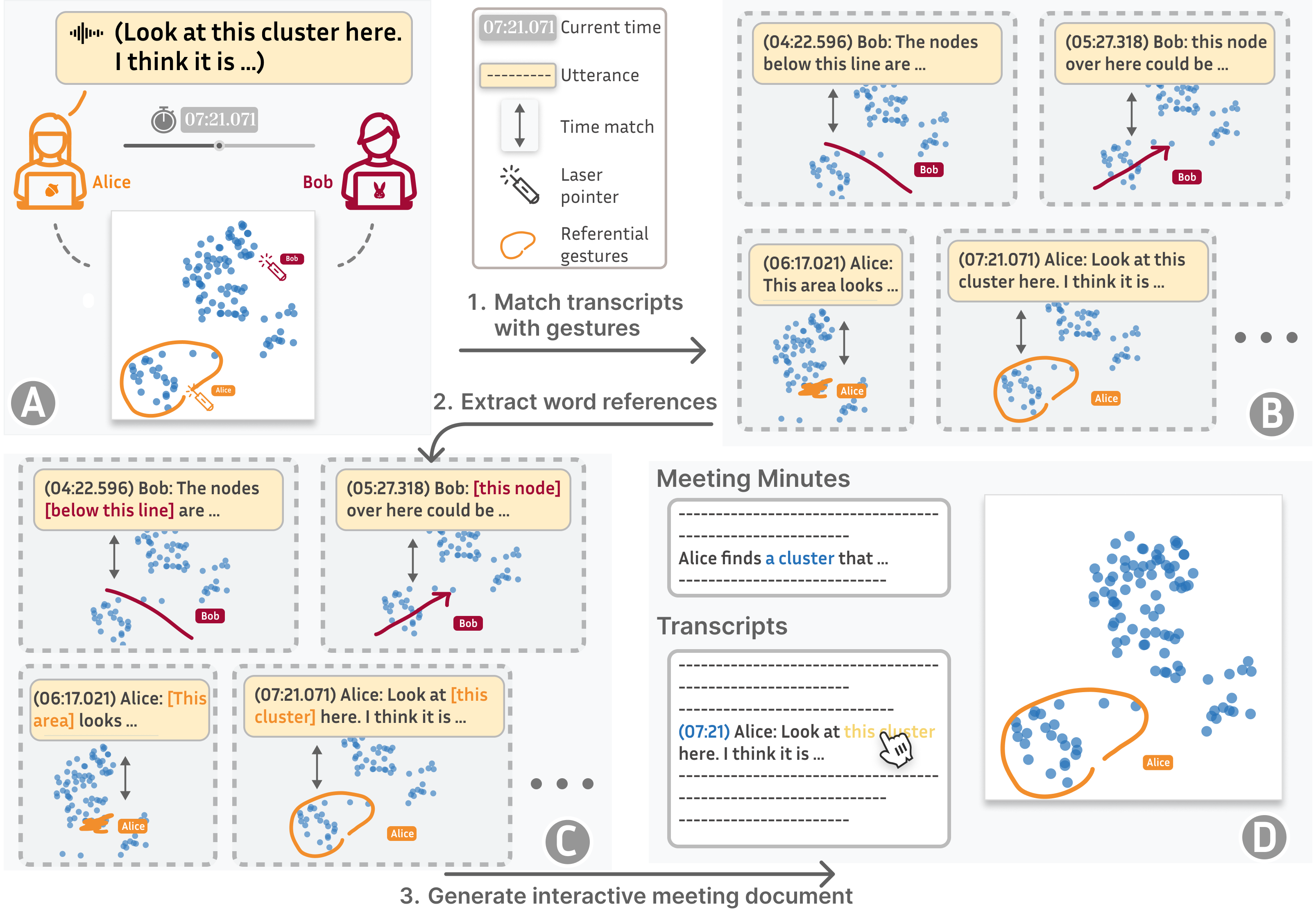}
  \caption{
  An overview of the proposed framework. It has four stages: (A) \textit{Data collection} that captures both deictic and verbal portions of the meetings. (B) \textit{Utterance matching} that associates referential gestures with the corresponding transcribed utterances. (C) \textit{Reference Extraction} that further extracts connections within the matched utterance. (D) \textit{Interactive Notes Generation} that represents collaborative data meetings with automatically generated transcripts and meeting minutes, both augmented with annotations.
   }
  \label{fig:teaser}
}

\graphicspath{{figs/}{figures/}{pictures/}{images/}{./}} 

\usepackage{pifont}
\usepackage{tabu}                      
\usepackage{booktabs}                  
\usepackage{lipsum}                    
\usepackage{mwe}                       
\usepackage{graphicx}
\usepackage{soul}
\usepackage{mathptmx}                  

\newcommand{\inlinehdr}[1]{\vspace{1ex}\noindent{\textbf{#1}}}
\newcommand\chang[1]{{\color{black}#1}}
\newcommand{\kate}[1]{\textcolor{black}{#1}}

\begin{document}


\firstsection{Introduction}

\maketitle

When people meet to present and discuss data and data visualizations, significant communication occurs through referential gestures such as pointing and indications of movement and flow~\cite{kita2003pointing}. These gestures enrich the {\em context} within which statements are made, playing a crucial role in shaping the meaning of those statements. This meaning-through-context is referred to as {\em deixis} by the linguistics community~\cite{stapleton2017deixis}. Among the various forms of communication, deixis holds a particularly important role in interchanges concerning visualizations~\cite{hill1991deixis}, providing clarity and depth to the explanation of insights. As shown in \autoref{fig:teaser}(A),  someone may verbally refer to ``this cluster'' while using a laser pointer to lasso points in the scatterplot for others to observe. In this example, verbal expression alone would fail to convey a clear interpretation without the accompanying referential gestures.

Despite being essential in communication around data, deixis is often overlooked when documenting collaborative visualization meetings. Traditional methods of documenting meetings each have their limitations: transcripts do not store referential gestures, while screen and video recordings may be missing pointer movements or lack fidelity to understand context from them. Additionally, manual note-taking can be distracting and must often be selective in what context it captures to keep up with the pace of the meeting~\cite{piolat2005cognitive}. Existing methods of documenting insights during visual analysis are either not designed for synchronous communication~\cite{viegas2007manyeyes, heer2007voyagers} or demand significant manual annotation and input~\cite{zhao2016annotation, chen2010click2annotate}, posing considerable challenges for their use in highly interactive settings like collaborative visualization meetings.

To facilitate future examination and review of communications around data, we propose a framework to document collaborative visualization meetings that underscores the importance of deixis.
We focus particularly on online meetings, where people synchronously discuss data visualizations in a distributed setting. In video conferencing, gestures to a digital visualization are often performed with the mouse or touch pointer. Video conferencing software~\cite{zoom2023, msteams2023} further allows people to annotate what is shown on screen with tools like the virtual laser pointer or pencil tool while interactive visualizations have deictic interactions such as brushing and highlighting~\cite{becker1987brushing, martin1995high}. We design our framework to capture these pointer-based deictic behaviors, storing them alongside the utterances.

Participants in virtual synchronous meetings {\em encode}, in the semiotics sense~\cite{hall1973encoding, chandler2022semiotics}, their deictic communication with a combination of audio and pointer gestures. To better understand how we can identify these communication pairs, we first conduct a formative study of the use of pointer-based gestures in online synchronous meetings around data. We use the findings to inform the design of our pipeline. We associate recorded gestures from annotation tools and visualization interactions with audio transcripts based on both temporal overlap and utterance semantics, utilizing the in-context learning capability of a large language model (LLM)~\cite{dong2022survey, min-etal-2022-rethinking} to perform the latter. With these matched gestures and dialogue, we generate an interactive document that provides both  meeting minutes, a compressed narrative of the meeting, and the full transcript, both linked with their matching referential gestures represented as animated annotations or interaction states.

To evaluate the effectiveness of our approach, we conducted a second user study involving our developed prototype, which is available on Github\footnote{\url{https://github.com/hconhisway/vitraexample}}. Overall, participants acknowledged the usefulness of our generated documents and expressed a preference for using them for meeting documentation over screen recording and audio transcription methods. Given that our study replicates gestural tools commonly used in collaborative meeting systems, we believe that the proposed concepts and design principles can be readily integrated into existing systems.

Our framework aims to {\em identify} deictic communication and re-encode it into documentation without decoding or changing the meaning, expecting the user to decode as they would have in the meeting. Recognizing the potential of decoding these deictic communications, we code the observed gestures from both of our studies and derive the first taxonomy for referential gestures in online synchronous communication. Though preliminary, this taxonomy can aid in understanding deictic pointing behaviors in online communication and provide a foundation for future projects that seek to recognize, capture, and design for such gestures.

In summary, our contributions are two fold:
 (1) A framework for generating deixis-informed interactive notes from synchronous online audio-visual meetings and
 (2) A preliminary taxonomy of pointer-based deictic gestures used in collaboratively exploring data visualizations.
\section{\kate{Background and} Related Work}
\label{sec:02-relatedwork}

We present relevant background in deixis, referential gestures, communication, and collaborative visualization. Then we discuss related work regarding insight documentation in visual analysis and methods for connecting text with visual components.

\subsection{Background on Deixis and Referential Gestures}
In linguistics, \textit{deixis} refers to the phenomenon where the meaning of certain words is dependent on the context in which they are used~\cite{stapleton2017deixis}. For example, the statement ``that apple tree over there'' cannot be fully understood without observing the speaker's index finger pointing towards the tree. Deixis can take on different forms, such as personal, spatial, and temporal. We focus on spatial deixis, where language is used to indicate location or direction within the speaker's contextual environment. We use the term \textit{referential gestures} interchangeably with deictic pointing behavior.

In the encoder/decoder model of Stuart Hall~\cite{hall1973encoding}, the communicator encodes their {\em meaning} as a message which is then {\em decoded} into a perceived meaning by the receiver. In our case, attention is directed through a combination of speech and gesture. We seek to identify and document both parts of the encoded message, essentially re-encoding the message for later decoding by the the meeting participants.

Previous work has underscored the critical role of referential gestures in human communication. Through surveying various types of communicative gestures, Clark~\cite{clark2003pointing} categorized them into two categories: pointing and placing. Pointing directs attention towards an object or location, while placing involves positioning objects in a way that communicates meaning. Clark argues these nonverbal actions are foundational to the way we convey and interpret meaning. Brennan et al.~\cite{brennan2005conversation} conducted an experiment with paired participants working together to analyze a map. By analyzing the movement of mouse cursors, they concluded that successful conversation depends on the dynamic interplay of both verbal and non-verbal (i.e., visual) cues. Building upon these works, our study acknowledges the critical role of non-verbal cues in communication. Particularly, we focused on the context of communication around visual representations.

Hill and Hollan~\cite{hill1991deixis} explored how deictic pointing behaviors encompass more than the simple directive of ``look here,''A discussing various hand gestures that convey diverse meanings. Heer et al.~\cite{heer2007design} further investigated the application of spatial referential gestures within the domain of asynchronous collaborative visualization. They distinguished between two primary types of referential gestures in spatial contexts. The first type encompasses brushing and dynamic querying, which are directly tied to data. They can support various automated tasks~\cite{yang2007analysis} and are applicable through different data views. The second type, \textit{graphical annotations} are more expressive, but are view-dependent due to their lack of data awareness. Building upon the taxonomy proposed by previous studies and our observations, we identify three types of referential gestures pertinent to remote synchronous communication around data visualizations: transient gestures expressed with a laser pen~\cite{oh2002laser, parker2005tractorbeam}, durable annotations made with a pencil tool~\cite{denisovich2005software, heer2007voyagers}, and manipulations of the visual interface activated by mouse actions~\cite{Fikkert2007}.

A critical issue related to referential gestures is ambiguity. In some scenarios, people may successfully communicate without using gestures. Clark et al.~\cite{clark1983common} demonstrated how ambiguity resolution is influenced by the familiarity among individuals. For instance, two individuals sharing a common understanding about certain flowers might effectively communicate by simply referring to ``this flower,'' while a third party, lacking this shared context, might be confused. 
The ambiguity problem also poses a threat to our documentation framework, as we cannot document information that is not explicitly expressed and relies on external context. We also discuss this limitation in Section~\ref{sec:09-discussion}. 

\subsection{Background on Collaborative visualization}
{\em Collaborative visualization} is defined as \textit{``the shared use of computer-supported, (interactive,) visual representations of data by more than one person with the common goal of contribution to joint information processing activities''}~\cite{isenberg2011collaborative}. Based on early works in the Computer-Supported Collaborative Work (CSCW) community~\cite{johansen1988groupware, applegate1991technology}, collaborative visualization scenarios are characterized by space (co-located vs. distributed) and time (asynchronous vs. synchronous) axes~\cite{isenberg2011collaborative, 5386648}. 
\chang{When designing our documentation framework, we focus on synchronous distributed collaborations, with an emphasis on the critical communication and discussion phases. Specifically, we envision small collaborative team meetings, described as ``Jam Sessions'' by Brehmer and Kosara~\cite{brehmer2021jam}, as primary use cases for our framework.}

\chang{Although our work does not aim to directly enhance collaborative visualization, we built the collaborative interface we use for data collection (Section~\ref{sec:04-collabinterface}) based on the previous efforts of collaborative visualization
~\cite{kim2010hugin, badam2014polychrome, schwab2021, badam2018vistrates}}. For example, 
VisConnect~\cite{schwab2021} introduces support for synchronizing low-level events in web-based interactive visualizations for collaboration. Neogy et al.~\cite{neogy2020representing} explored the design space of interfaces for remote synchronous collaborative visualizations, with the aim of enabling effective collaboration among users while preserving the autonomy necessary for independent work. \kate{We use these works to inform our testbed so that our framework operates on ecologically valid scenarios.}

\subsection{\kate{Other Related Work}}

\chang{\inlinehdr{Insight documentation in visual analysis.}} \chang{Previous research has recognized the important role of notes made during visual analysis~\cite{mahyar2012note},} though not in the context of remote synchronous collaborative visualization. In these other contexts, observed benefits included that analysts can review insights previously recorded and use the records as materials for deeper analysis and organization~\cite{mahyar2012note, mahyar2014supporting}. \chang{Apart from taking notes in separate medium, annotating visualizations can also be considered a form of knowledge externalization and can also facilitate the analysis process~\cite{zhao2016annotation}.} Kim et al.~\cite{kim2019inking} found that free-form annotations, as opposed to note-taking in a separate medium, may encourage participants to maintain a state of ``flow,'' which is associated with heightened creativity. Walny et al.~\cite{walny2017active} also used a free-form pen in their study to ask participants to draw on visualizations while reading. Their results suggest that active reading behaviors transfer from documents to visualizations. We draw an analogy between our provided pencil tool \chang{(shown in Section~\ref{sec:04-collabinterface})} and the free-form pen in non-collaborative scenarios. 
\chang{Our framework also aims to document insights from visualizations. We focus on remote synchronous meetings as the primary use case, prioritizing the automatic generation of notes, as manual note-taking can be challenging in such highly interactive scenarios.}

\label{sec:0204-interactivedocs}
\chang{\inlinehdr{Connecting text to visual media.}} In the digital era, many documents produced and consumed online incorporate both textual and visual elements. 
Many works have sought to harness computational technologies to enhance the readability and user experience of online document text.
For example, Kong et al.~\cite{kong2014extracting} introduced a crowdsourcing pipeline for extracting references between texts and charts and an interactive application that highlights these correspondences on selection. Kim et al.~\cite{kim2018facilitating} developed an interactive document reader that automatically links document text with corresponding spreadsheet cells. This linking was shown to enhance users' ability to match text with tables.
\chang{
Badam et al.\cite{badam2018elastic} further introduced a contextual visualization technique that can automatically link text contents with tables and provide visualizations based on the reader's current focus. Kori\cite{latif2021kori} presented a mixed-initiative approach for building references between charts and text through both manual input and algorithmic suggestions. EmphasisChecker~\cite{kim2023emphasischecker} can detect and highlight mismatches in emphasis between line charts and textual descriptions. In addition to utilizing visualization to enhance the reading experience, textual descriptions (annotations) are also employed to augment visualizations. Lai et al.~\cite{lai2020automatic} developed a pipeline that uses computer vision techniques to automatically annotate visualizations based on provided textual descriptions.} Similarly, ChartText~\cite{pinheiro2022charttext} links text with statistical charts. They discussed a use case of automatically annotating charts to a presenter's description during a video conference as motivation, but did not demonstrate it in practice.
\chang{
Our work resonates with previous efforts to connect text to visual media, but targets a different type of reference between text and media: those encoded by participants in virtual meetings, expressed through a combination of deictic references and spoken words.}
\section{Formative Study \& Design Rationale}
\label{sec:0601-formativestudy}
To inform our design, we conducted a formative study of how people use referential gestures in remote synchronous meetings around visualizations. We first designed and implemented a collaborative interface for use in this study, \kate{representing such intefaces in general}, based on a survey of existing collaborative working tools. Then, we recruited four participants with expertise in the design and analysis of visualizations.
Participants were asked to share and discuss their own visualization projects as well as visualizations they found engaging through the collaborative interface. These visualizations included common charts (e.g., Gantt charts, line charts, scatter plots), scientific visualizations (including volume rendering results), and visual analytics systems. We detail the findings from the observations and discussions, and explain how they influenced our design decisions.
We first describe our collaborative interface. Then we present the findings of our formative study and the design guidance we interpreted from them.

\subsection{\kate{Representative} Collaborative Interface}
\label{sec:04-collabinterface}

In designing our \kate{representative} interface to collect deictic gestures, we encountered limited discussion and guidance regarding how they are expressed in remote environments. 
Early works focused on the use of the mouse \kate{pointer} to direct viewer's intentions~\cite{brennan2005conversation,hill1991deixis}.
However, the volume of data generated by the trajectory of mouse movements is substantial and does not always indicate a referential gesture, limiting its capacity for preserving records.

We surveyed how referential gestures are expressed in popular collaborative working tools, including videoconferencing tools, such as Zoom~\cite{zoom2023} and Microsoft Teams~\cite{msteams2023}, 
and whiteboard tools, such as Mural~\cite{mural} and Excalidraw~\cite{excalidraw}. We found that referential gestures are often expressed using a laser pointer metaphor. 
Compared to the mouse \kate{pointer}, the virtual laser pointer is more expressive as it maintains temporary trails that can express deixis more naturally with diverse meanings beyond just ``look here.''
In addition to the laser pointer, we also found these applications have a free-drawing pencil tool to create durable annotations in collaborative visual analysis that last until deliberately erased, such as, for example, marking a location as ``Area A'' and referring to it afterward. We thus chose to implement these two common tools in our interface.

\begin{figure}[!tb]
    \centering{\includegraphics[width = 1\linewidth, alt={The image is a full screen capture of our collaborative interface client. There are three major sections. The left, annotated as 'A,' has a grid of image thumbnails with a plus button to add more. The middle, annotated as 'B,' shows an example visualization showing colored points where different regions are mostly the same color. This is a common visualization showing a multivariate projection with clusters assigned. Above the visualization is a tool bar with icons showing a cursor, a pencil, an erasor, and a laser pointer. The laser pointer is highlighted, indicating it is the active tool. The tool bar has a sub-label, 'b1.' Over the visualization there is a wobbly yellow circle around one of the clusters with the tag label 'Alice' in the same yellow color indicating that Alice made the gesture. This annotation is labeled 'b2.' The right portion of the interface, labeled 'C', has several additional items. The top shows participants as circles of their own color and initial with a yellow circle with an 'A' for Alice. This participant view is labeled 'c1.' Underneath is a toggle control labeled 'Interactive Demo Off' which is labeled 'c2.' Underneath that is a button that says 'Recording... ' next to a chart showing successive bar heights from left to right which indicate the user's voice level. This indicator is labeled 'c3.'}]
    {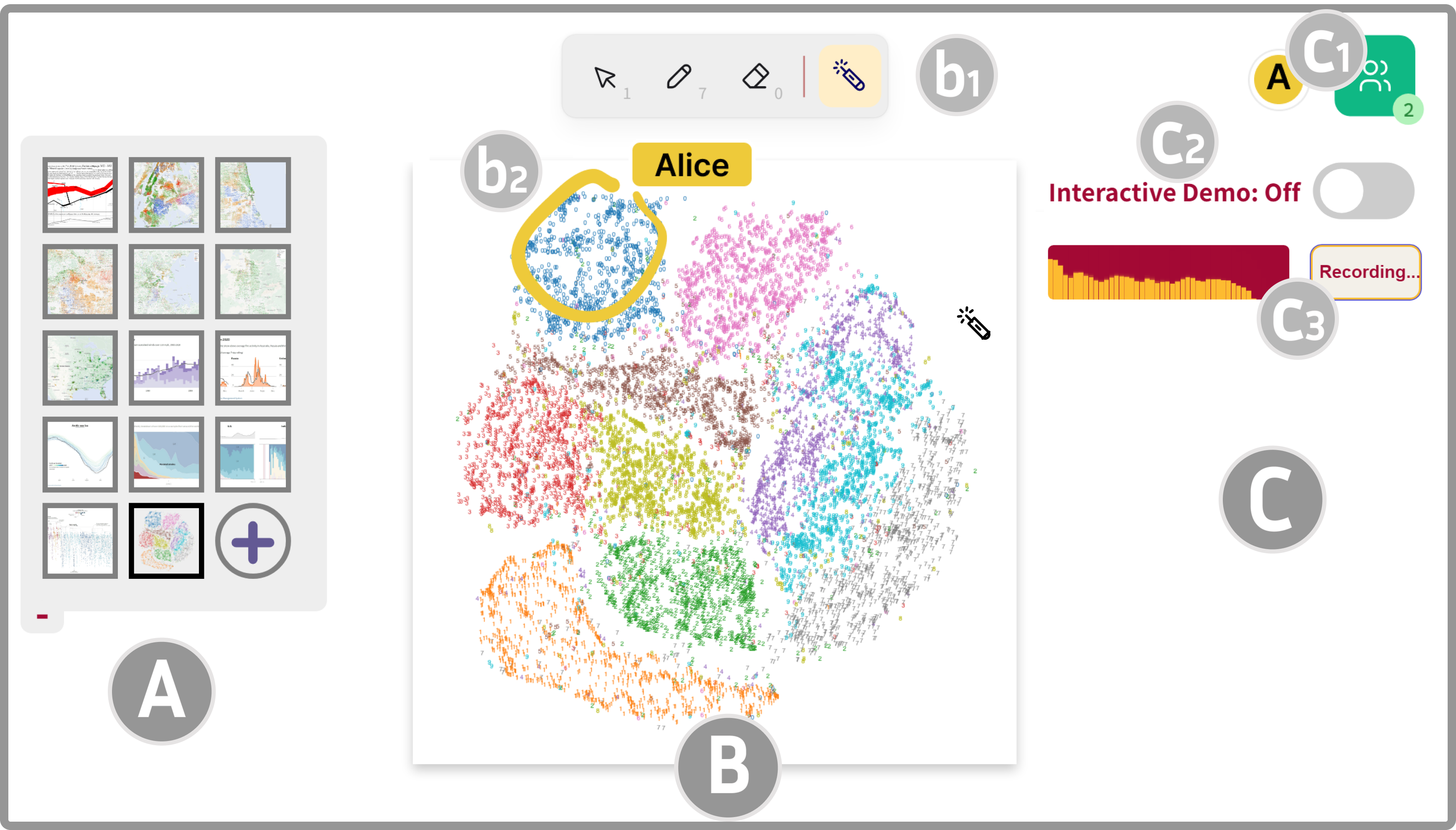}}
    \caption{An overview of the collaborative interface, serving to enable collaborative visualization and data collections. It consists of (A) a gallery of meeting materials, (B) a collaborative board, and (C) room controls.}
    \label{fig:interface}
\end{figure}

\subsubsection{Collaborative Interface for Capturing Gestures}
Based on the above investigation, we designed a collaborative interface (\autoref{fig:interface}). 
It consists of three main parts: (A) the gallery of meeting materials , (B) the collaborative board, and (C)) the room controls.

\inlinehdr{(A) Gallery of meeting materials.} Anticipating multiple visualizations may be examined in a meeting, we designed a gallery of meeting materials. This gallery is synchronized among all users and acts as a menu for selecting the main visualization being viewed. The gallery concept draws a parallel to the slide gallery in presentation tools, but is designed to offer greater flexibility in a more informal structure. It is intended for housing raw materials that require analysis and discussion within the meeting context.

The gallery supports direct import of static visualizations in JPEG, PNG, and SVG formats. As the collaborative interface's support for interactive web-based visualizations requires some manual effort to work with our system, we do not yet support automatic upload. Other types, such as videos, we expect will require other design considerations for appropriately capturing referential gestures. 

\inlinehdr{(B) Collaborative board.}
The collaborative board is the main area of the collaborative interface and where we expect in-depth analysis of visualizations and the referential gestures of the participants to occur. We base our design on the Excaldraw whiteboard tool~\cite{excalidraw}.

We incorporate the virtual laser pointer and pencil tools in the toolbar at the top of the screen (\autoref{fig:interface}$b_1$). From left to right, the toolbar includes the mouse pointer for initiating interactions (further details in Section \ref{sec:0402-ProvenanceTracking}), the pencil tool for durable annotations, an eraser for removing them, and the laser pointer for transient gestures.
\autoref{fig:interface}$b_2$ shows an example of Alice, a user remote to the viewer, using the laser pointer to circle a cluster in the scatterplot.
The yellow trail is synchronized with the view of the viewer (local user). It remains visible for a brief period before gradually fading away.

\inlinehdr{(C) Room controls.} We enable users to initiate a new session by starting a `room' and sharing its link with others (\autoref{fig:interface}$c_1$). This feature is also inspired by Exalidraw. It includes a real-time display of participants currently active within the room. These controls also include the audio recording functionality, as indicated by the button in \autoref{fig:interface}$c_3$, which is essential for the data collection process.

\subsubsection{Supporting Interactive Visualizations}
\label{sec:0402-ProvenanceTracking}

\begin{figure}[!tb]
    \centering{\includegraphics[width = 1\linewidth, alt={The image illustrates a client-server interaction for a network visualization application where nodes and edges are represented as a node link diagram, on the right there is a bar chart showing some statistics, each bar represents a node in the graph. In the scenario depicted, Client 1 displays a network graph with various nodes and edges, with a specific node labeled "Marius" highlighted and a hand cursor shown clicking on it. This click event on the node "Marius" triggers an interaction, resulting in a highlight in the correspoinding bar chart. The server receives this click event from Client 1 and stores the state, which includes information about the selected node ("Marius") and the positions of the nodes. This state information is then stored in a database. Subsequently, the server sends the click event to other clients, causing them to update their views to reflect the selection of the node "Marius."}]
    {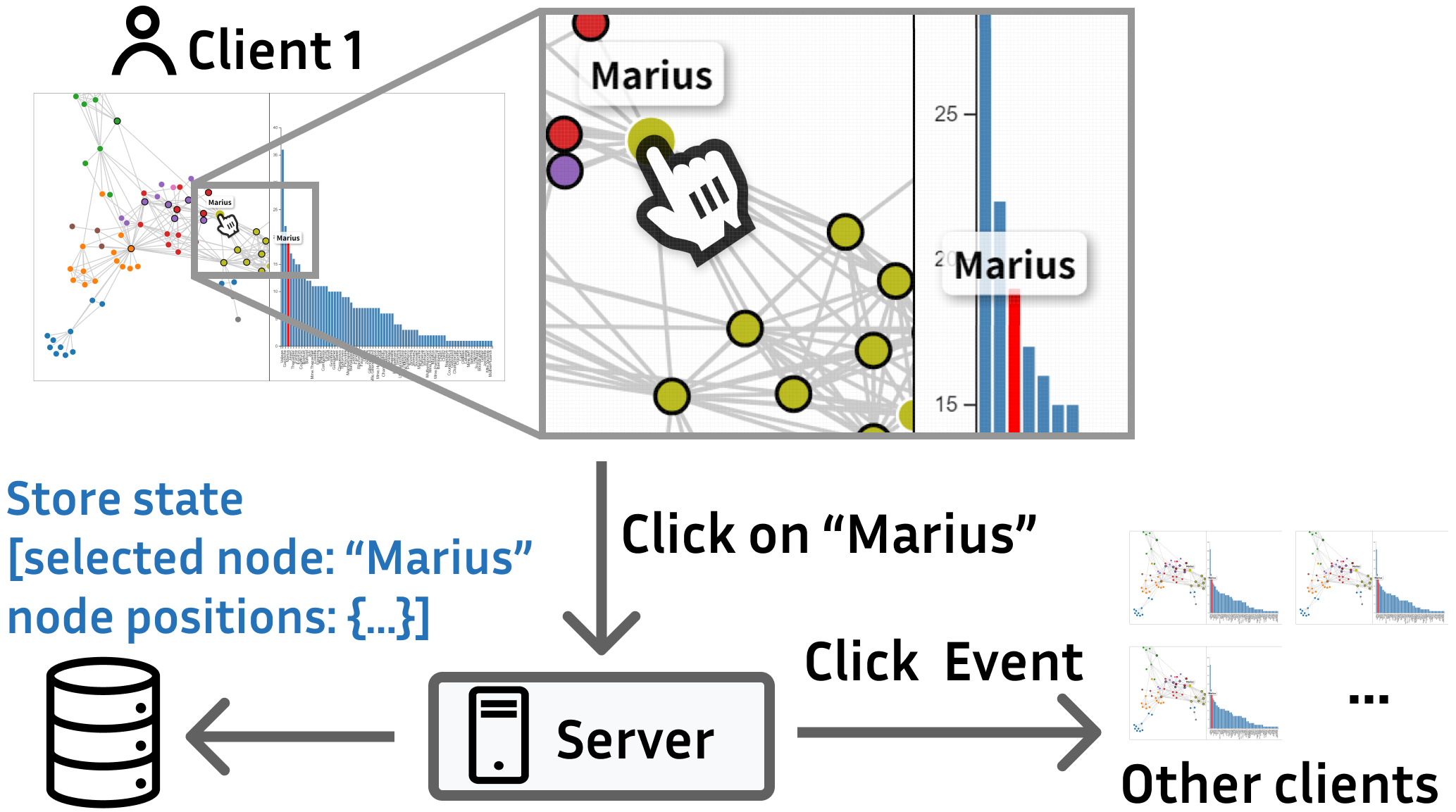}}
    \caption{An illustration of the operational mechanisms underpinning the interactive visualization demo within our collaborative interface. In this example, ``Client 1'' clicked on the node ``Marius''. This event is relayed to the server and then broadcast to all connected clients. Concurrently, the change of the current selected node and node positions are stored in the state file.}
    \vspace{-2mm}
    \label{fig:interactiveexample}
\end{figure}

We developed a proof-of-concept interactive visualization within our interface to demonstrate our deixis-centered documentation for remote synchronous meetings in interactive scenarios. 
As illustrated in \autoref{fig:interactiveexample}, we employ the event synchronization mechanisms provided by VisConnect~\cite{schwab2021} to ensure that changes within the visualization are coherently synchronized across different clients. Moreover, each user interaction with the visualization triggers the recording of the current interaction state. These interactions represent another form of deixis.

Currently, there is no existing solution that can import interactive visualizations directly into the collaborative interface without some coding effort.
While VisConnect enables collaborative visualization through event synchronization, it still
requires some implementation changes to the non-collaborative interactive visualizations.

Capturing the deictic actions on the interactive visualization requires provenance tracking, which has been extensively researched in the visualization community~\cite{xu2020survey, ragan2015characterizing, north2011analytic}.
Recently, tools like SIMProv~\cite{camisetty2018enhancing} and Trrack~\cite{cutler2020trrack} have emerged, offering capabilities for capturing provenance in web-based interactive visualizations.
We employ trrack~\cite{cutler2020trrack} to collect provenance data related to user interactions. 
Trrack can record the state of the interactive visualization at each interaction event, enabling the reconstruction of the visualization's state at any point in the user's journey. 
This feature is helpful for generating interactive notes that accurately reflect the current visualization state.

\subsection{Formative Study Observations \& Outcomes}

We make the following observations based on the activities of the participants during our formative study:

\inlinehdr{Observation: The laser pointer and pencil tool engendered different gesture types, frequency of use, and degrees of precision.}
As the laser pointer was transient and the pencil tool durable, they were used in different ways by participants. First, the laser pointer was primarily used to direct attention, often accompanied by verbal cues such as ``look here.'' In contrast, the pencil marks were made to last a long time. 
One participant noted that they would opt for the pencil tool to highlight discussion points they deemed important. So durable annotations made with pencil tool often span a large portion of a discussion while gestures made with laser pointer are typically associated with one sentence. 

Second, the laser pointer was used significantly more than the pencil tool. Two participants mentioned they preferred staying with the laser pointer and only switched to the pencil tool when necessary. They would switch back to the laser pointer after completing their tasks with the pencil. This behavior underscores the laser pointer's role as the primary tool for most users. 

Third, referential gestures with the laser pointer tended to lack precision, \chang{which means drawings may not precisely cover intended area,} and rely heavily on verbal clarification to resolve.
For example, a participant might use the laser pointer to indicate a growing segment in a line chart, saying, ``look at this growing period...'' The line drawn might not exactly cover the intended segment, so the verbal explanation of ``growing period'' helped convey the intended meaning. In contrast, annotations made with the pencil tool were characterized by greater precision. Participants tended to be more cautious with the pencil, sometimes using the eraser to correct inaccurate annotations.

\textbf{Design Outcome:} Given the vastly different natures of how the transient laser pointer and durable pencil tool are used, we developed distinct strategies for linking them with text in interactive documents. We choose to link transient gestures in a fine-grained fashion, like phrases or keywords. We choose to treat durable annotations more like image changes in the interactive document.

These choices influence our technical requirements, particularly in the utterance matching and reference extraction steps described in Section~\ref{sec:06-meetingnote}. For transient gestures, we match utterances to the laser pointer's actions based on the timestamps captured at the start (mouse down) and end (mouse up) of its activation. In contrast, durable annotations are linked to every sentence articulated from their creation (indicated by the pencil tool's mouse down action) until their deletion (erased with the eraser). 

Furthermore, due to the imprecise nature of transient gestures, we implement an additional reference extraction step. This process aims to connect these gestures with specific words and phrases in the corresponding sentence, thereby creating more intuitive and clearer linkages.

\inlinehdr{Observation: People use the laser pointer for purposes beyond gesturing.}
The use of referential gestures among participants showed significant variation. Some individuals used the laser pointer beyond gesturing, occasionally creating drawings that were meaningless. One participant characterized these actions as involuntary and an unconscious byproduct of their thought process. Generally, these behaviors do not hinder interpersonal communication, as others tend to disregard these non-essential gestures. However, these indiscriminate doodles can introduce inaccuracies in the resulting interactive document by generating irrelevant annotations. 

\textbf{Design Outcome: } To mitigate the inclusion of spurious or non-deictic gestures in our interactive notes, we take additional measures to identify and filter out extraneous gestures.

\inlinehdr{Observation: The motion of the laser pointer is used to attract attention, indicate direction, and convey emotional intensity.}
The motion of the laser pointer can convey additional meaning. First, motion can be used to better direct attention. Especially when referring to small areas, people tended to use the laser pointer to draw back and forth over the same location. Second, motion can convey critical information such as reading direction. For example, a person can draw a line to guide viewers to follow a path from a certain direction. In the absence of animation, the indication of direction becomes ambiguous. Third, motion can be used to convey a sense of emotional intensity. For example, by rapidly accelerating the movement of the pointer to highlight a rapidly growing area on a line chart. 

\textbf{Design Outcome:} In light of these insights, we adapted our interface to record the timestamps of each point along the laser pointer's trails. This data allows us to recreate the referential gestures through animations, accurately mirroring the original motions.

\section{Automatic Generation of Interactive Notes}
\label{sec:06-meetingnote}

We present our framework for automatically generating interactive meeting notes with both verbal and non-verbal communication cues. We first present a brief overview. We then go over the techniques used to transform the recorded audio and gestures into interactive notes.

\subsection{Framework Overview}
\label{sec:0601-frameworkoverview}
The goal of our framework is to enhance documentation of collaborative meetings around visualization. The framework comprises four main components: i) data collection (\autoref{fig:teaser}A), ii) utterance matching (\autoref{fig:teaser}B), iii) reference extraction (\autoref{fig:teaser}C), and iv) interactive notes generation (\autoref{fig:teaser}D). \kate{The first these steps identify messages with deixis-encoded aspects. The final step re-encodes the identified message as a persistent digital document.} \autoref{fig:teaser} illustrates this workflow. We describe the framework in overview here and present details in the subsequent sections.

\inlinehdr{i) Data Collection.} To accurately document both deictic and verbal portions of collaborative visualization meetings, we collect three types of data: audio recordings, referential gestures (using \chang{pointer}-based laser pointer and pencil tool), and interaction provenance data. Each data type is labeled with timestamps to facilitate subsequent matching processes. We use the collaborative interface from our formative study, though the framework could be applied to any interface that collects this data. 
In our implementation, the recorded audio is transcribed using Whisper~\cite{radford2023robust}, an open-source automatic speech recognition (ASR) system. Whisper further provides word-level precision timestamp prediction, which helps in the later utterance matching step. We also utilize the implementation of speaker diarization using Whisper~\cite{Diarization} to distinguish different speakers.

\inlinehdr{ii) Utterance Matching.} Once data is collected, the audio is transcribed and the transcribed utterances are aligned with corresponding referential gestures based on timestamps.
For example, in \autoref{fig:teaser}B, the transcribed utterance ``Alice: Look at this cluster here...'' is matched with the referential gesture of Alice circling a cluster in the scatterplot using the timestamps of the gesture and speech.
Drawing from our formative study with visualization experts, we developed specific matching strategies for different types of referential gestures. These strategies are discussed in detail below.

\inlinehdr{iii) Reference Extraction.} We further winnow the audio associated with a gesture to a more precise meaning using a large language model (LLM). We design prompts 
to extract connections between words or phrases within the matched utterances and referential gesture pairs.
For example, in \autoref{fig:teaser}C, the LLM identifies ``this cluster'' from the transcribed utterance ``Alice: Look at this cluster here...'' as the object most likely to be the focus of the gesture.
The design and implementation of this reference extraction process are described in Section \ref{sec:06-meetingnote} \kate{and the} practical limitations are discussed in Section \ref{sec:09-discussion}. \kate{We note this step partially decodes meaning through the LLM to refine the identification. We hypothesize further understanding of deixis encoding could improve this and other applications and thus present a preliminary taxonomy of pointer-based deictic gestures in Section \ref{sec:08-taxonomy}.}

\inlinehdr{iv) Interactive Notes Generation.} We create an interactive document representing the collaborative data meeting with \chang{meeting minutes}, a transcript, and annotated visualizations. The \chang{meeting minutes} are generated by the LLM. Both the \chang{meeting minutes} and transcripts are augmented by creating links in the text to animated annotations that display the referential gesture in the visualization.
For example, in \autoref{fig:teaser}D, hovering over the word ``cluster'' in the interactive note triggers a replay of the referential gesture made by Alice---the circle she made around the cluster replays in the scatterplot. We describe the design and implementation of interactive notes generation below.

\subsection{Gesture Filtering \& Matching}

We have different strategies for matching transient, durable, and interactive visualization gestures to sentence-sized utterances. However, first we filter out easily identifiable non-referential gestures. As noted in Section~\ref{sec:0601-formativestudy}, some people tended to doodle when not speaking. Keeping only gestures made by the active speaker is a simple yet effective strategy to filter out most non-communicative gestures. 

\begin{figure}[!tb]
    \centering{\includegraphics[width = 1\linewidth, alt={
    The image is a composite chart displaying the stock prices of four companies (Apple - AAPL, Amazon - AMZN, IBM - IBM, and Microsoft - MSFT) over time from 2000 to 2010. The chart includes annotations labeled A, B, and C, with associated commentary.
    A: "IBM's stock price was the highest at the beginning." (Green highlight on IBM's stock price at the start)
    B: "Look, here, here, and here, it seems that an economic crisis has caused all stock prices to fall." (Blue circles on specific points in time across all stock price charts)
    C: "Apple's stock had a prolonged downturn starting in 2000...", "...reached its peak here before 2008...", "...and then a trough appeared..." (Purple annotations on Apple’s stock price chart indicating specific trends)
    The stock price charts show:
    AAPL: A prolonged downturn starting around 2000, peaking before 2008, and then experiencing a trough.
    AMZN: Relatively stable with some fluctuations and significant peaks.
    IBM: Initially the highest stock price among the four companies, with notable peaks and troughs.
    MSFT: Generally stable with minor fluctuations over the period.
    }]
    {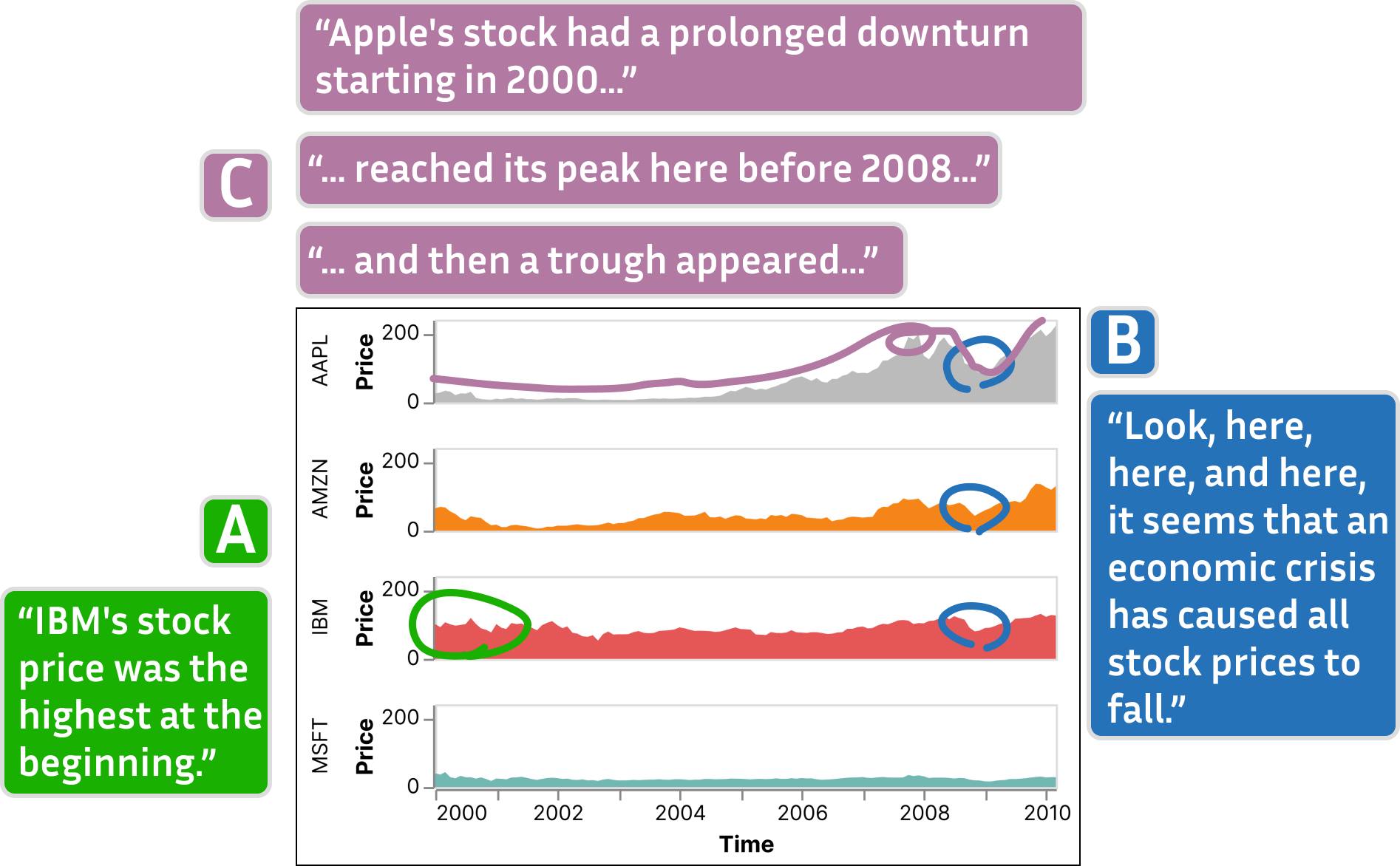}}
    \caption{\chang{Illustration of various scenarios} for matching utterances with transient referential gestures. (A) One gesture matches with one sentence. (B) Several gestures match with one sentence. (C) One gesture matches with several sentences.}
    \vspace{-2mm}
    \label{fig:diffLaser}
\end{figure}

\inlinehdr{Transient gestures.} Referential gestures made with the laser pointer are the most common. They typically have a short duration, usually with one gesture corresponding to a single sentence. However, this heuristic is not sufficient. During our experiment, we also observed many cases where people used multiple gestures within a single sentence. Drawing a long-lasting trajectory while speaking multiple sentences was a very rare case. ~\autoref{fig:diffLaser} illustrates these three cases. In case \includegraphics[width=0.27cm]{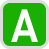} and case \includegraphics[width=0.27cm]{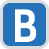}, where a gesture is associated with at most one sentence, it is sufficient to record the match by assigning each gesture to the single sentence it appears during. 

Conversely, scenarios like \includegraphics[width=0.27cm]{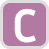}, where multiple sentences are linked to a single gesture,  typically arise when the speaker uses the laser pointer continuously without releasing the mouse button. In such instances, a single gesture may convey different meanings at different junctures, thus presenting a challenge to the following reference extraction task. This complexity arises because extracting meaningful references from the gesture becomes more difficult as the size of the text (utterance) grows. To mitigate this issue, we divide the continuous gesture into distinct segments based on the timestamps associated with each sentence. This process ensures that each gesture is paired with a single sentence at most.

\inlinehdr{Durable gestures.} As discussed in Section \ref{sec:0601-formativestudy}, durable annotations made with the pencil tool persist from the creation to the deletion. Rather than associate these with all of the sentences uttered during their existence, we instead associate them with timestamps. The intent is to match them with a state of the visualization rather than specific utterances. Thus, in the output interactive notes document, users can reveal durable annotations by timestamp. \autoref{fig:interactivenote} shows these timestamps in the interactive notes, enclosed by red rectangles.

\inlinehdr{Interactive visualization gestures.} Multiple factors could be considered when handling stored manipulations of visual interface (interaction provenance). These factors include the presence of animated transitions, the duration of the transitions, the type of interaction, and the extent of changes made to the visualization. Some interactions, like selection, highlighting, and brushing, are analogous to the mouse \chang{pointer} gestures in terms of focusing attention. We decided not to consider animated transitions as these are properties of the interactive visualizations rather than a choice of the person manipulating the system. \kate{It would be unclear whether the animated transition is an intended part of the encoding of the speaker or an aspect they could not control and therefore we rely on the instigating action, e.g., selection or brushing, instead.} We also chose not to consider interactions that alter the visual encoding, as those are more akin to looking at another visualization rather than gesturing. We chose to include dynamic queries that do not change the visual encoding as gestures for their attention and focus aspects. Given the durability of the changes, we do timestamp matching instead of utterance matching, similar to durable gestures. We further discuss the nuance of different interaction types and possible ways to capture them in (interactive) documentation in Section~\ref{sec:09-discussion}.

\subsection{Reference Extraction}
\begin{figure}[!tb]
    \centering{\includegraphics[width = 1\linewidth, alt={
    This is an image with two parts, the left part points to the right part through an arrow, under the arrow is the text "LLM". The two parts are almost identical but with different colors and annotations. On top of the left part is the text, says "“Since its foundation, after several years of rapid growth, we can observe that Google's stock price reached its peak around 2008." The whole text is highlighted in yellow with a cursor hand pointing on it. Below the text is a line chart of stock prices. There are annotations on the line chart, shown as two circles circling around one part of rapid growth and another part of a peak.
    The right part also has the same text on the top, but only "several years of rapid growth" and "reached its peak around 2008" are highlighted, the cursor hand is pointing at the "several years of rapid growth". On the line chart below, there is only one annotation annotating the part of rapid growth.
    }]
    {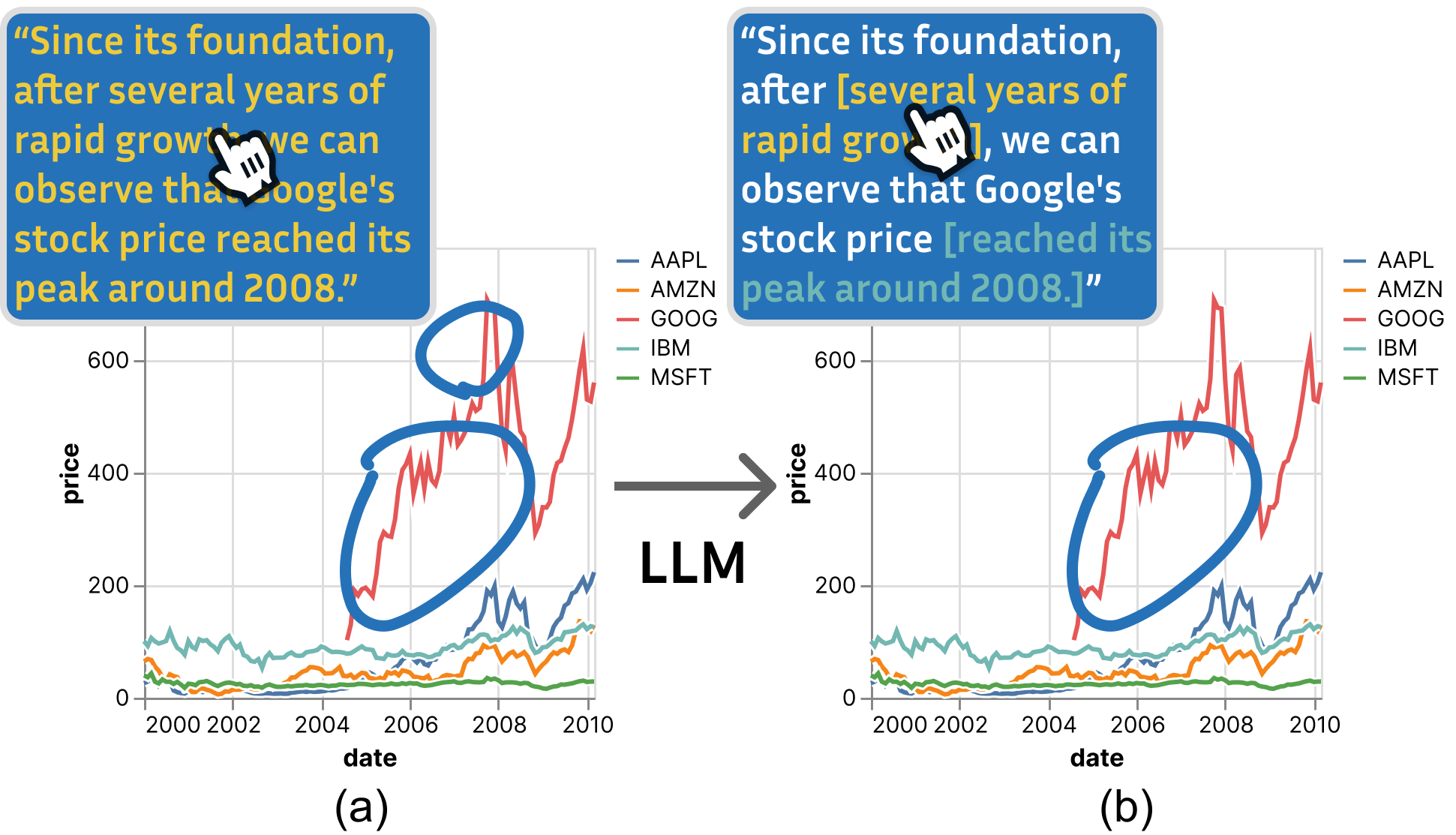}}
    \caption{Illustration of reference extraction: (a) Two imprecise gestures are matched with one sentence. (b) After reference extraction, gestures are sequentially associated with two separate phrases.}
    \vspace{-3mm}
    \label{fig:reference}
\end{figure}

The casual and imprecise characteristics of transient gestures, which we discovered in our formative study, pose additional challenges to associating those gestures with the most meaningful speech. We cannot typically discern the annotator's intent based on the gesture alone. Reviewing the speech (text) is necessary to grasp the intended meaning fully. This issue is compounded in instances where multiple gestures are made within a single sentence, as demonstrated in \autoref{fig:reference}(a).
To make the generated notes more comprehensible and reduce the cognitive load on users, the reference extraction step matches gestures to relevant words and phrases from their previously matched sentences. As illustrated in \autoref{fig:reference}(b), two gestures are identified and sequentially associated with ``several years of rapid growth'' and ``reached its peak around 2008'', thereby making both user interaction and reading clearer and easier to understand.

We employ the in-context learning capabilities of LLMs~\cite{dong2022survey, min-etal-2022-rethinking} for reference extraction. In-context learning allows the LLM to adapt to new tasks through inference alone,  eliminating the need for additional training or fine-tuning. To enhance the precision and reliability of the extraction process, we adopt the ``chain-of-thought''~\cite{wei2022chain} prompting strategy coupled with a few-shot learning approach~\cite{brown2020language}. This approach can enhance the model's robustness by explicitly demonstrating the thought process required for the task, encouraging the model to follow a step-by-step cognitive approach. The ``chain-of-thought'' prompting, in particular, aids in making the model's decision-making process more transparent and interpretable. The few-shot examples in the prompt can be found in the supplemental materials.

\subsection{Design and Implementation of Interactive Meeting Note}
\begin{figure*}[!tb]
    \centering{\includegraphics[width = 1\linewidth, alt={
    This is a figure showing the interaction note application. On the right there is a screenshot of a webpage. The webpage is vertically splitted by a line into two parts. The left part shows texts of meeting minutes of meeting from 00:14:45 to 00:16:24, the right part shows a map of segregations in Boston, on top of the map is a gallery of several visualizations used in the meeting. The texts in left part has several parts highlighted in blue. In the screenshot, the cursor is currently pointing to one phrase hightlighted in blue "red dots scattered throught the area". At the same time, the map in the right shows several hand-drawn circles circling around several area in the city. There are labels "Bob" around these circles meaning that they are drawn by bob.
    The left part is text showing the corresponding transcript of the meeting minutes. Arrows shown nine different highlighted phrases in the transcripts are transformed into the three highlighted phrases in the meeting minutes. Highlights numbered "1,2,3" in transcripts are merged into the one highlight in meeting minutes, while highlights numbered "4" and "9" are seperated kept by the meeting minutes. On top of the transcripts, there is a note saying that "Note: This illustration shows a pair of corresponding transcripts and meeting minutes. On the actual page, they are arranged vertically and can be navigated by scrolling, 
    as shown in Figure 1 (D)."
    }]
    {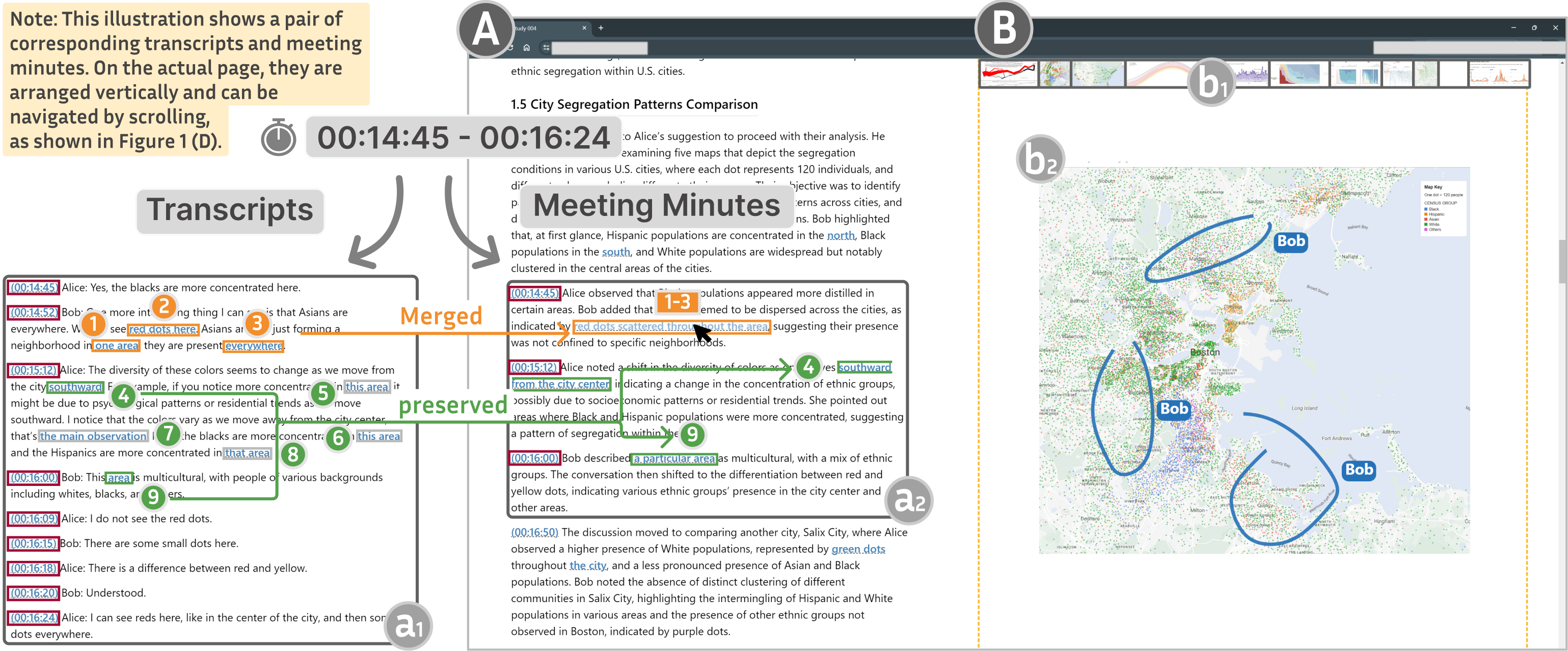}}
    \caption{An overview of the interactive notes, with: (A) Interactive text, comprising transcripts from audio and the LLM-generated \chang{meeting minutes}, includes interactive text components based on the results of utterance matching and reference extraction. (B) Visual media from the meetings are presented with annotations based on parameters transmitted by the interactive text on the left. This operation can change the underlying visualization, add annotations, and alter interactive states.}
    \vspace{-2mm}
    \label{fig:interactivenote}
\end{figure*}

We design the interactive note document following the design outcomes we identified during the formative study (Section \ref{sec:0601-formativestudy}). As shown in \autoref{fig:interactivenote}, the document consists of two primary views: the \textit{Transcripts \& \chang{Minutes} view} (\autoref{fig:interactivenote}A) and the the \textit{Gallery \& Visualization view} (\autoref{fig:interactivenote}B).

The Transcripts \& \chang{Minutes} view depicts the verbal communication of the meeting. The transcript is generated directly from the audio as described in Section \ref{sec:0601-frameworkoverview} i). However, raw transcripts can be lengthy, unstructured, and noisy, making it challenging to quickly identify key points and actionable items within the text. Recently, researchers have explored how to leverage an LLM's in-context learning capacity to understand complex, nuanced meeting content effectively~\cite{schneider2023team, rousseau2023darbarer}. Following a similar strategy to Schneider et al.~\cite{schneider2023team}, we produce \chang{meeting minutes (long-form summaries)} by segmenting meetings into topics and generating \chang{meeting minutes} for each topic separately. This method can produce \chang{meeting minutes} that are cost-effective compared to using large language models across the whole transcript.

During \chang{the generation of meeting minutes}, we apply an additional step to instruct the LLM to preserve referential gestures when \chang{transforming} transcripts, merging multiple gestures when necessary. As shown in \autoref{fig:interactivenote}, $a_1$ and $a_2$,  five referential gestures were preserved in the \chang{meeting minutes} out of nine total gestures in the transcripts. \includegraphics[width=0.32cm]{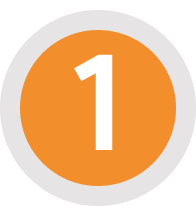}, \includegraphics[width=0.32cm]{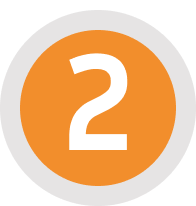}, and \includegraphics[width=0.32cm]{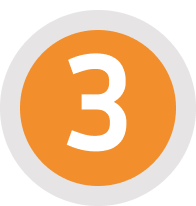} all depicted the Asian communities in Boston, and are semantically merged into \includegraphics[width=0.52cm]{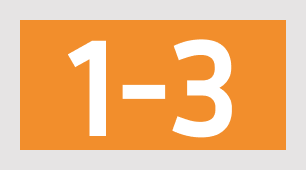} in the \chang{meeting minutes}. When a mouse hovers over \includegraphics[width=0.52cm]{13.png}, 
the visualization view (\autoref{fig:interactivenote} $b_2$) displays the three gestures made by \textit{Bob}, which are a combination of gestures \includegraphics[width=0.32cm]{1.png}, \includegraphics[width=0.32cm]{2.png}, and \includegraphics[width=0.32cm]{3.png}.
Of the next five gestures, \includegraphics[width=0.32cm]{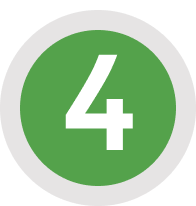} and \includegraphics[width=0.32cm]{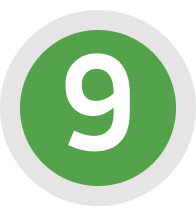} are preserved and included in the \chang{meeting minutes}, while the intervening gestures are only available in the transcript.

We build our interactive notes using prior work on authoring interactive documents~\cite{conlen2018idyll, heer2023living}. Specifically, we \kate{use} the interaction features of Living Papers~\cite{heer2023living}, a framework for integrating executable code, interactive components, and traditional text into a unified document. We write the reference extraction results into Living Papers' custom markdown. 
When the mouse hovers over the \kate{text links (light blue)}, as shown in~\autoref{fig:interactivenote}(A), the parameters \chang{(data regarding the referential gestures)} are passed to the \textit{Gallery \& Visualization view} (\autoref{fig:interactivenote}B), which then switches between different charts and plays the annotations based on the provided parameters. This approach provides the additional feature that users can edit the interactive notes directly through this markdown file.

\section{User Evaluation}
\label{sec:07-userevaluation}

We conducted an evaluation with users to assess the utility of our approach and to further examine the use of gestures in remote synchronous meetings with data visualizations. \kate{The study was approved by the University of Utah Institutional Review Board (00175137).}

\subsection{Study Design}

Our study was conducted in one-hour sessions with paired individuals who collaboratively discussed visualizations through our system and then reviewed the generated documents. While we expect collaboration dynamics may be different with larger groups, we chose to conduct the evaluation in pairs as we were focusing on the gesture-capture aspects of our framework. We consider our sessions as a simplified version of small collaborative team meetings described by Brehmer and Kosara~\cite{brehmer2021jam}. In these ``Jam Sessions'' people meet and discuss the data and data visualizations casually.

\subsubsection{Participants}
We recruited 18 participants (10 men, 8 women, all fluent English speakers) to evaluate our note generating tool. \chang{Based on our recruitment channels,} all participants are graduate students in biology or computer science \chang{from a university in North America}. Participants were matched into pairs based on their availability. Participants were compensated 30 USD for their time.

We sought participants that would be likely to perform exploratory data analysis. As such, we advertised to the following groups: (1) a data science course at our institution, (2) a data science meetup at a second institution, (3) a visualization course at a third institution, and (4) the graduate student group at our institution.

\subsubsection{Procedure}

We began our sessions with briefing and consenting. The facilitator (first author) then gave a demonstration of our collaborative UI and its features ($\sim$10 minutes). Participants were then asked to turn on their audio recordings, view the visualizations already in the interface, and discuss them with each other based on prompt analysis questions in the interface. ($\sim$30 minutes). The visualizations and prompts are available in supplemental materials.  

After the analysis session, the interactive documents were then generated and provided to the participants. Before reviewing the documents, the participants were asked to fill out a background survey about their experiences with collaborative data meetings and note-taking. The facilitator then verbally verified with each participant that the linked gesture annotations were available to them. \chang{The facilitator then asked the participants} to review the documents and fill out a second post-survey (described below). This process took about 20 minutes. Participants were then debriefed and the session ended.

\subsubsection{Post-Survey}

The post survey consisted of seven 5-point scale questions and three open response questions. It was administered through Google Forms. We asked participants to (1) rate the quality of the interactive documents and their specific components, (2) compare the documents to video recording and transcripts, and (3) rate their likelihood of using a framework like ours in the future. Each of these sections was followed by an open response asking them to elaborate. The full survey with responses is available in the supplemental material.

\subsubsection{Visualizations Used in the Study}
We chose a variety of visual idioms for the participants to explore during the session \kate{as} different idioms \kate{might} encourage different gestures. 
The static visualizations included the Minard map of Napoleon's march, segregation maps of several cities from the New York Times~\cite{nytimes2015census}, and (stacked) line charts, node-link diagrams with topics of global warming and virus evolution from Reuters~\cite{reuters2020coronavirus, reuters2022weather}. The interactive visualization is a Les Misérables co-occurrence graph with a bar chart displaying the frequency of occurrence for each character's name, shown in \autoref{fig:interactiveexample}. We asked participants to select the visualization that they deemed most likely to spark discussion. Each visualization was discussed by at least two groups.

\subsection{Results}

We present the results of both post-surveys.

\subsubsection{Data Meeting Background Survey Results}
In our background survey, we asked the participants to report their experiences with collaborative data meetings and note-taking. We discuss the findings from this survey. The raw survey results are available in supplemental materials.

Most participants (17/18) have experience with data analysis, of which 13 reported having done so with visual aids. Six participants claimed that they engage in meetings that are similar to our setting (collaborative data meeting with visual aids) frequently at school or during work.

When asked about their experiences with note-taking, the majority (11/18) reported jotting down some key points from their meetings. Three participants reported they do not take notes during meetings, with two adding they do not have time during the meeting. Three participants said they generate notes after meetings by watching video recordings or just using their memory. Fourteen participants said they use digital or physical writing tools (e.g., iPad, pen and paper) to take notes. Eight participants use a text editor and keyboard input (e.g., Notepad and Visual Studio Code).

We then asked the participants how they incorporate visual data into their notes. Eleven participants said they would put screenshots in their notes. Nine participants \kate{said} they would draw a sketch. Five participants \kate{said} they would write text description of the visual items. 

Lastly, we asked participants to discuss the main challenge they face in taking and using notes in meetings with data analysis. Eight participants (P3, P5, P7, P8, P9, P10, P12, P17) mentioned time constraints and the fact that talking while writing can be distracting. Seven participants (P1, P4, P6, P9, P12, P13, P16, P18) wrote about the difficulty of relating data (visualizations) with text. These responses validate the motivation for our framework.

\begin{figure}[!tb]
    \centering{\includegraphics[width = 1\linewidth, alt={%
      This is a table showing results. The first three rows are ratings from 1 (Poor) to 5 (Excellent). Overall quality has 2 with ratings of 3, 8 with ratings of 4, and 8 with ratings of 5. Summary quality has 4 with ratings of 3, 6 with ratings of 4, and 8 with ratings of 5. Transcript quality has 3 with ratings of 3, 5 with ratings of 4, and ten with ratings of 5. The next four are on a scale from Strongly Disagree to Strongly Agree. The first category Preference over video recordings, which has 1 Disagree, 1 Neutral, 6 Agree, and 10 Strongly Agree. The second is Preference over transcripts with 1 Disagree, 7 Agree, and 10 Strongly Agree. The third is Would use for multi-session work, with 1 Disagree, 7 Agree, and 10 Strongly Agree. The last is Would use for note-taking with 1 Disagree, 4 Agree, and 13 Strongly Agree.
    }]
    {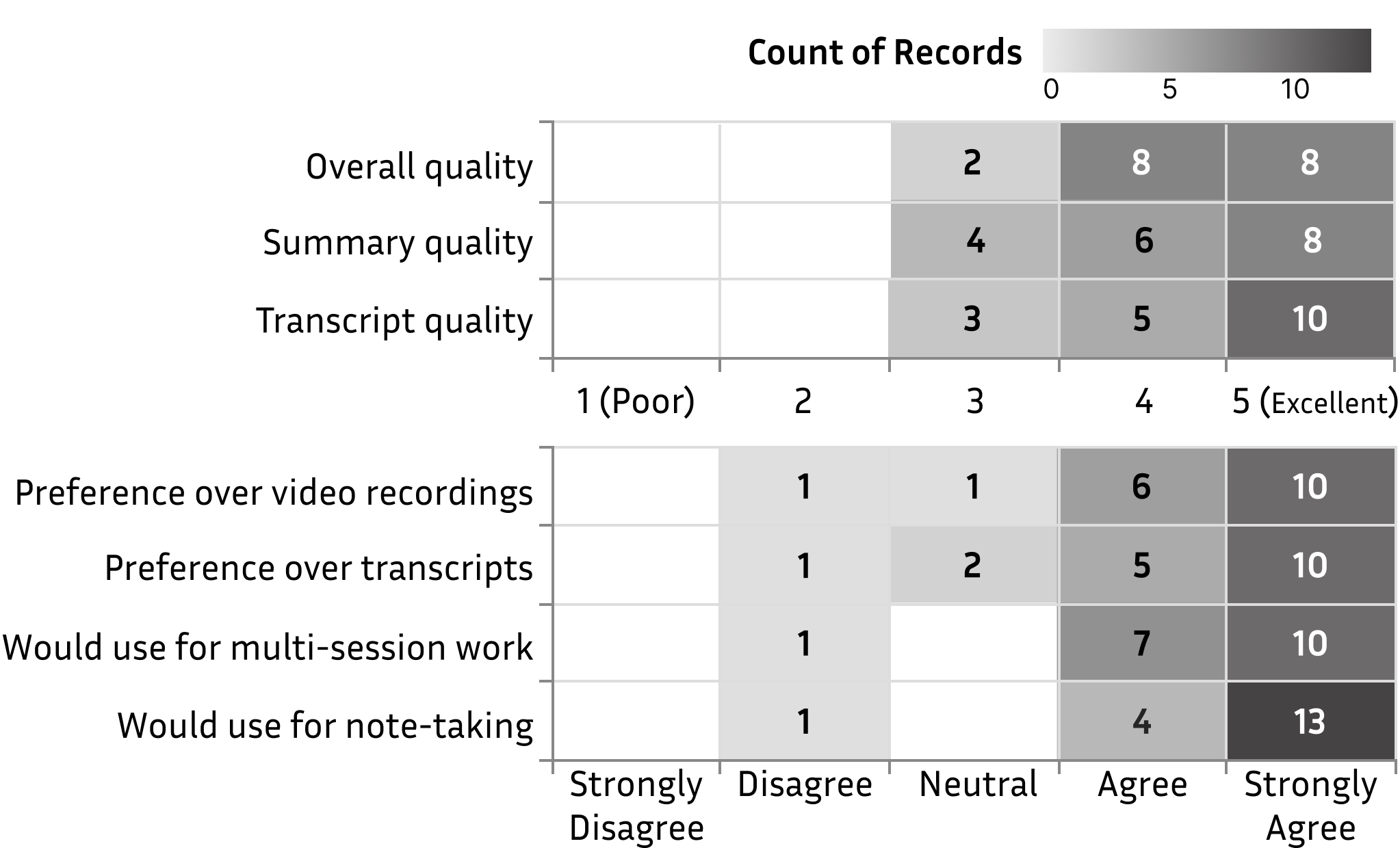}}
    \caption{Participant responses to scale-based survey questions. Most participants preferred our interactive notes to video recordings and transcripts and would use them again, despite mixed responses to the quality.}
    \label{fig:responseResults}
\end{figure}

\subsubsection{Interactive Notes Evaluation Survey Results}

\autoref{fig:responseResults} shows the results from our scale-based questions. We note that most participants preferred the interactive notes over reviewing video or the transcript alone and most would use a tool like our framework for both reviewing during a multi-session project or documenting a collaborative online meeting. Participants were more mixed on the quality of the framework. We look at the open responses to better interpret these results.
The second author coded the open responses and coalesced them into the insights below:

\inlinehdr{Participants noticed errors in both the transcript and \chang{meeting minutes}, but many considered both good enough despite these errors.} Ten participants \kate{(P2, P4-5, P7, P9-10, P14, P16-18)} wrote \kate{that there were} errors in either the transcript, \kate{specifically incorrect or dropped words}, or \chang{ the meeting minutes}, \kate{specifically incorrect or missing details. F}our \kate{(P5, P10, P14, P18)} also mentioned the notes were good despite these errors. Three participants \kate{(P5, P7, P14)} additionally noted what they considered matching errors in the gesture correspondences.

\inlinehdr{Most participants found utility in the interactive documents, citing reasons like the speed of reviewing the notes and labor saved during meetings.} Nine participants \kate{(P1-3, P5-6, P16-18) suggested the} interactive documents were better than video, with \kate{all but P17} alluding to the fact that they could read and seek much faster than in video. \kate{P17} noted \kate{instead that} the context is clearer through interactive documents:

\begin{quote}
The advantage of the tool is that it has annotated visualizations, which makes it a lot easier to get the context. I guess you can still achieve that with a screen recording to some extent but you may need to hit pause a couple of times. -P17
\end{quote}

Eight participants \kate{(P1, P3, P6, P8, P13, P15-16, P18)} picked out the value of the \chang{meeting minutes} as useful for situations where they did not need specific details. Three participants \kate{(P1, P13, P18)} specifically mentioned the annotations.

Seven participants \kate{(P2-3, P5, P10, P13, P15, P17)} mentioned the framework is labor-saving: they did not have to take notes themselves or summarize them after the fact. For some, this would allow them to focus more on the meeting:

\begin{quote}
I do collaborative meetings and having this tool during my discussion session would be helpful. It means I wouldn't have to concentrate about taking notes. Instead I can focus fully on the discussion with my collaborators. -P13
\end{quote}

\inlinehdr{Some participants noted video has advantages of higher accuracy and the inclusion of human expressions.} When comparing with video,
\kate{three participants} (P4, P14, P18) noted the video would not have errors. Two (P8, P16) suggested \kate{the video may contain precise details the automated system might note capture.} \kate{Two participants} (P7, P12) noted that video captured human cues from voice and expression \kate{where text notes do not, suggesting these cues enhance understanding.} \kate{Additionally, P14 wrote they were used to searching videos from their coursework and thus would prefer videos over transcripts.}

\inlinehdr{Results Summary.} Overall participants generally found utility in the framework, with many wanting to use it again and noting its features over other automatic methods such as video recordings and transcripts. Views on transcription and subsequent summary quality were mixed, which limited the utility of the framework.

\subsection{Limitations}

We did not pair participants based on their mutual familiarity. However, previous work in non-\kate{pointer}-based settings has found that the familiarity between two people can affect the referential gestures they use~\cite{clark1983common}. People with greater familiarity may successfully communicate using fewer gestures. Whether this holds true for online meetings with \kate{pointer}-based gestures remains to be investigated.

We chose to limit the evaluation sessions to collaborative pairs instead of larger groups. While we did not expect this choice to have an \kate{large} effect on overall transcription quality or the kinds of gestures used, \kate{we do not know the extent to which it may have affected utterances, repetitions, or audio collisions and our framework's subsequent performance}. Other studies are needed to further explore the use of \kate{pointer}-based gestures in remote synchronous multi-person meetings around data.

The facilitator knew seven of the participants before they volunteered as they work in the same building. Those participants may have been more generous in their responses.
\kate{Participants acquainted with the facilitator rated an average quality of 4.43/4.14/4.43 (overall, summary, transcript), in contrast to the total averages of 4.33/4.32/4.39.}

\section{Preliminary Taxonomy of \kate{Pointer}-based Deictic (Referential) Gestures}
\label{sec:08-taxonomy}

In our participant \kate{studies}, we observed that gestures using the laser pointer and pencil were more expressive compared to those made with the mouse \kate{pointer} alone. Participants made a variety of deictic gestures beyond just directing attention. We thus sought to better understand the nuances of these behaviors to better recognize, process, \kate{potentially decode}, and design for these gestures in this and future tools. 

The first author reviewed all study recordings to identify each gesture and assess its intention. From this review, we found patterns in the types of gestures used and their purpose.
Such patterns underscore the complexity and richness of non-verbal communication facilitated by digital tools, thus providing a foundation for advancing our understanding of non-verbal communication through digital media.

Following the previous efforts of categorizing hand gestures~\cite{hill1991deixis} and the exploration of referential behavior in asynchronous communication~\cite{heer2007design}, we describe the first preliminary taxonomy for referential gestures based on simulated laser pointers. 
We classify the gestures based on user intentions, emphasizing the critical notion that the significance of these gestures is deeply rooted in the context of communication.

\begin{figure}[!tb]
    \centering{\includegraphics[width = 1\linewidth, alt={%
    The first part of this figure is a table with rows for each intention and columns for different gestures found. The bottom is text descriptions. Together they say: A - Direct Attention with examples of a Z/N shape over the target, an arrow pointing to a target, a circle around a target, and a wavy line underneath a target. B is Highlight Trends with a line along a line chart, a circle in the peak of a line chart, and a wavy line to indicate rise and fall. C is Depict Path which has an example of a laser pointer following a path and a circle over a path in the visualization. D is Outline Boundary which has an example where the pointer traced an existing boundary in an image, a straight line that divides space, and more complex lines to divide space. E is Indicate Area/Group which has a lasso-selection of a cluster and also a Z/N shape covering a cluster. F is Refer to Absect Objects. It has drawings of a missing part of a contour and a circle of a missing part of the data. G is Indicate Interval. It has two circles on the ends of an interval, a line indicating the span of an interval, and a circle over an entire interval. H is Connecte Components and shows a line between compoents in one example and multiple circles indicating component members in another. I is Direct Reading Attention. It has a line suggesting reading order in one example an arcs to indicate reading direction in another.
    The second part of this figure is more detailed explanation of the first part. For each column in first part, it explained several examples of how this would work.
    A. Direct Attention
        1. Z/N-shaped scanning at target.
        2. Draw an arrow to point to the target.
        3. Circle the target with the laser pointer.
        4. Draw wavy lines under the target.
    B. Highlight Trends
        1. Trace trends along a line chart.
        2. Highlight a pivot as a narrative anchor, e.g., 'after this point'.
        3. Draw a polyline in a blank space to convey intent, such as 'rising and then falling'.
    C. Depict Path
        1. Laser pointer moves along the path.
        2. Draw circles on the path to denote it, suitable for charts with fewer paths.
    D. Outline Boundary
        1. Move the laser pointer along the target's outer boundary, may slow or pause at certain spots for emphasis.
        2. Draw a line to differentiate content on both sides.
        3. Use more complex shapes to differentiate multiple areas.
    E. Indicate Area/Group
        1. Lasso-select the area/group to be highlighted.
        2. Z-shaped or N-shaped scanning to denote area/group, works when the group boundary is obvious or combined with verbal description.
    F. Refer to Absent Objects
        1. Redraw the contour of the missing part, e.g., draw a line with arrow to denote a missing axis.
        2. Circle the location of the missing part.
    G. Indicate Interval
        1. Direct attention to both ends of an interval to indicate its range.
        2. Mark interval with line segments.
        3. Encircle the interval with a circular frame, supplemented by verbal description.
    H. Connect Components
        1. Draw a line connecting two components.
        2. Continuously direct attention across multiple objects to indicate connection.
    I. Direct Reading Direction
        1. Draw a straight line in a specific direction to indicate reading order.
        2. Use arcs or other shapes to denote more complex reading directions.
    }]
    {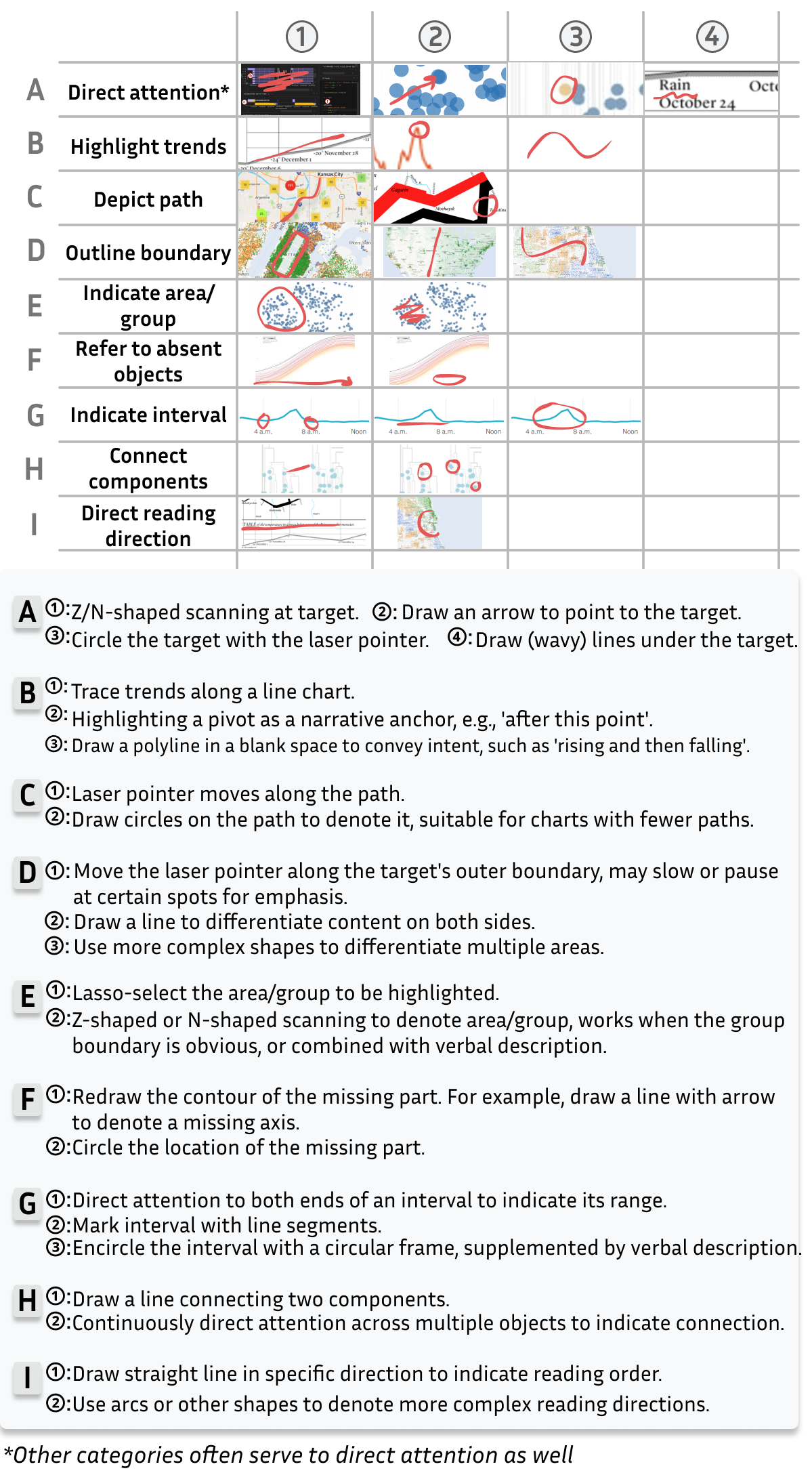}}
    \caption{Our preliminary taxonomy derived from observations made during our two studies. Letters A to I correspond to different intentions, with each intention associated with multiple distinct gestures, denoted by \ding{172}-\ding{175}. Text descriptions of each gesture are in the key in the lower half.}
    \vspace{-3mm}
    \label{fig:taxonomy}
\end{figure}

Our taxonomy categorizes the observed gestures into nine intentions, each served by multiple gestures. \autoref{fig:taxonomy} lists these intentions and the gestures observed. While each observed intention-gesture pair is supported by numerous examples from our study, \kate{our dataset is limited in scale and scope and thus this taxonomy should be considered {\em preliminary}.It can be used to inform additional, more comprehensive studies. In assembling this preliminary taxonomy, we observed:} 

\inlinehdr{i) Intentions can be served by multiple distinct gestures.} In \autoref{fig:taxonomy}, we note that each intention, from A to I, can be served by multiple different gestures. Besides personal preference, this appears to be related to the type and attributes of the targeted object. For example, when trying to direct attention, people tend to draw wavy lines under text (A\ding{175}), use Z/N shape scanning for large objects (A\ding{172}), and employ arrows to point to smaller objects (A\ding{173}). This finding underscores the inherent diversity and adaptability in non-verbal communication, providing insights for understanding deictic behavior in visualization context. It also suggests that intention might be derivable when taking mark type and other visual features into account.

\inlinehdr{ii) Gestures often convey dual meanings.}
On one hand, they all serve to direct attention, guiding the gaze of other participants. 
On the other hand, many of them (all categories except A, which is just ``direct attention'') serve to convey their own specific meanings, such as highlighting trends (B) or indicating intervals (G). This demonstrates the rich expressiveness inherent in the deictic gestures.

\inlinehdr{iii) The meaning of referential gestures relies heavily on the accompanied verbal expression.}
From the table, it is evident that many gestures employed to convey various intentions share fundamentally similar forms, such as A\ding{174}, B\ding{173}, and C\ding{173} all appearing in the form of a small circle. This reveals another critical characteristic of referential gestures: their meaning is contingent upon the accompanying verbal expressions, underscoring the inseparable relationship between these non-verbal cues and verbal expressions.

\section{Discussion and Future Work}
\label{sec:09-discussion}

\kate{We proposed} a new framework for capturing referential gestures together with utterances \kate{to document} collaborative data meetings. To the best of our knowledge, it is the first tool that tries to document meetings following a deixis-centered approach. As shown in the evaluation, the utility of the generated notes was recognized by most users. Compared to traditional recording methods such as screen recording and audio transcription, our approach was more preferred.

The interactive notes generated by our framework resemble the interactive documents discussed in Section~\ref{sec:0204-interactivedocs}. Our work differs from these methods primarily because it does not seek automated extraction of references, but instead capitalizes on natural human input—referential gestures used in communication—to obtain the references between texts and charts. 

Our approach tends to result in fewer annotations than fully automated methods like ChartText~\cite{pinheiro2022charttext}, as gestures are not constantly used with every verbal expression. In many scenarios, the speakers do not need to use the laser pointer to direct people’s attention. For example, they can just use the labeled names or colors to refer to a certain entity and the audience will be able to quickly identify the referred target. Speech without gestures \kate{frequently} occur when the visualization is less information-rich or when the cognition process is pre-attentive (such as color pop-out).

\kate{We designed our system to only have annotations in deictic situations to preserve the original modes of directing attention, such as when pure verbal cues utilize salient labels or features alone.}

However, this does not account for the impact of ambiguity in communication. The lack of referential gestures could also be due to a high degree of familiarity and common ground regarding the topic between the two parties. For such cases, further investigation is needed.

Our work also has limitations. First, the quality of the reference extraction step could be improved. It is limited by both the in-context learning capabilities of the LLM and situations where a single utterance can contain multiple potential objects that could be referenced by gestures, making it difficult to determine the intended referent. Solving this problem requires improved techniques to further extract information from gestures and the visualization. The multimodal capabilities of an LLM may potentially resolve this issue, but at present, such improvements would require substantial computational resources. An alternative approach would be to refine and extend the taxonomy we have provided\kate{, thereby enabling further decoding of meaning}. By combining contextual information about the target visualization, such as chart and mark types, with their accompanied utterances, we \kate{might} better infer the type of gesture, thus providing more reliable links. As such, more data regarding \chang{pointer-based} gestures in collaborative settings is needed.

Another limitation is that our framework only considers a limited set of visualization interactions that can serve as referential gestures. We implemented a highlight interaction in our proof-of-concept demonstration, as such interactions are clearly analogous to other deictic gestures. However, interactive visualizations in the wild typically involve multiple different types of interactions. Some interactions, such as highlighting and querying, do not alter the encoding method of the visualization, while others may change the visualization entirely, such as transitioning from a node-link diagram to an adjacency matrix. This type of interaction is distinct from deixis and more akin to image switching in static visualization discussions. Further study is needed to understand the intersection of deixis and interactive visual manipulations. 

\chang{Our framework focuses on the automated documentation of collaborative meetings around visualization. Compared to manual note-taking, automated documentation allows participants to focus on the meeting with fewer distractions, a point several participants brought up in our evaluation.
However, research has shown that structured note-taking can benefit cognitive engagement in online classes~\cite{fang2022understanding}.
Our design goal was to aid more interactive settings where  participants frequently communicate and discuss with each other. The trade-offs in using such an automated approach versus manual note-taking in a more one-way setting such as online classes would require further investigation.}

\acknowledgments{%
We thank the study participants, anonymous reviewers, Md Dilshadur Rahman, Zach Culter, Alexander Lex, Kiran Gadhave, Connor Scully-Allison, Sayef Sakin, Shadmaan Hye, Kalina Borkiewicz, and Utah SCI members for their valuable feedback.

The work reported here was supported by the Defense Advanced Research Projects Agency (DARPA), under agreement HR00112290092. The views and conclusions contained herein are those of the authors and should not be interpreted as necessarily representing the official policies or endorsements, either expressed or implied, of the Defense Advanced Research Projects Agency (DARPA) or the U.S. Government.}

\bibliographystyle{abbrv-doi-hyperref}

\bibliography{vitranote}








\end{document}